\documentclass[aps,floats,epsf, twocolumn,showpacs]{revtex4}
\usepackage{graphicx}
\def\be{\begin{equation}}
\def\ee{\end{equation}}
\def\bea{\begin{eqnarray}}
\def\eea{\end{eqnarray}}
\def\bdm{\begin{displaymath}}
\def\edm{\end{displaymath}}
\def\ba{\begin{array}}
\def\ea{\end{array}}

\begin{document}

\title{Pairing symmetry and vortex zero-mode for superconducting Dirac fermions}


\author{Chi-Ken~Lu}
\author{Igor F.~Herbut}
\affiliation{Department of Physics, Simon Fraser University, 8888
University Drive, Burnaby, B.C. V5A 1S6, Canada}


\begin{abstract}

We study vortex zero-energy bound states in presence of pairing
between low-energy Dirac fermions on the surface of a topological
insulator. The pairing symmetries considered include the $s$-wave,
$p$-wave, and, in particular, the mixed-parity symmetry, which
arises in
 absence of the inversion symmetry on the surface. The
zero-mode is analyzed within the  generalized Jackiw-Rossi-Dirac
Hamiltonian that contains a momentum-dependent mass-term, and
includes the effects of the electromagnetic gauge field and the Zeeman coupling as
well. At a finite chemical potential, as long as the spectrum without the vortex is fully gapped,
the presence of a single Fermi surface with a definite helicity always leads to one Majorana
zero-mode, in which both electron's spin projections participate. In particular, the
critical effects of the Zeeman coupling on the zero-mode are discussed.




\end{abstract}

\pacs{03.75.Lm,74.20.Rp,74.45.+c}

\maketitle

\section{Introduction}

The zero-energy Majorana modes have recently become a topic of
many investigations in condensed matter physics, perhaps due
to the concomitant non-Abelian statistics\cite{ReadGreen} and
potential applications to fault-tolerant quantum
computation\cite{nayak}.  In a weakly-coupled triplet $p$-wave
superconductor, for example, one such zero-mode was found
in the core of a half-vortex, and two pairs of such vortices
are required to perform the non-Abelian braiding
operations\cite{Ivanov}. The vortex bound states were, of course, already studied a
long ago by Caroli, de Gennes, and Matricon in a $s$-wave
weak-coupling superconductor,\cite{deGennes} but a pair of true
zero-modes was discovered only much later, by Kopnin and Salomaa in a triplet
superfluid\cite{KopninSalomma}. Both studies were performed within
the framework of Bogoliubov-de Gennes (BdG) equations which
describe  the low-energy electron-hole excitations around the
\emph{degenerate} Fermi surfaces in the presence of
superconducting instability. The degenerate Fermi surfaces can be
split,  and the superconducting order parameter can contain both
the spin singlet and triplet components when the inversion
symmetry is broken\cite{Gorkov}. For example, the corresponding
vortex bound states in a noncentrosymmetric
superconductor,\cite{Lu2}  or the Andreev bound states on the
domain wall\cite{Tanaka} have also been studied. The existence of
such zero-modes has been shown to depend on a Z$_2$ topological
invariant associated the mixed order
parameter\cite{Fujimoto,ZhangTopology,YipTopology}.

The metallic surface of a topological insulator (TI) resembles the
noncentrosymmetric material, as in both cases the Fermi
surfaces are chiral, i.e. the projection of the spin is locked to
the direction of the momentum. However, an important  difference is
that the surface of the topological insulator has a only one such  Fermi surface
 \cite{TIrev}. Recently, an interesting system consisting of
this peculiar surface subjected to the superconducting proximity
effect has been proposed to bind one Majorana zero-mode inside  a
$s$-wave vortex core of unit vorticity\cite{FuKane}. This
zero-mode is nothing but the mid-gap state of the
Jackiw-Rossi-Dirac Hamiltonian\cite{Jackiw1} (JRD), studied before
in the context of charge fractionalization  in
polyacetylene\cite{SSH} and graphene,\cite{HCM,Jackiw2} with its
existence assured by an index theorem\cite{IndexW}. In addition,
the zero-modes of the JRD Hamiltonian have also been studied
recently in the context of the insulating
\cite{HCM,IgorInsulating,IgorDirac} and superconducting states on
the graphene's honeycomb
lattice\cite{WilczekSC,Kopnin,Babak,Bergerman,Igor2010}.

In this paper we study the existence and the form of the
zero-modes of a class of the general JRD Hamiltonians describing
paired Dirac fermions in presence of a superconducting vortex in
the spin-triplet $p$-wave state, and in the mixed singlet-triplet
state, when the inversion symmetry is absent. We pay special attention to the
effects of a finite chemical potential\cite{Kopnin,Bergerman} and
the magnetic fields on the zero mode, which are all present in
real systems. The problem under study is also relevant to graphene,
where  different symmetries of the superconducting phases have
also been considered\cite{CastroNeto,Annica,Roy}. In graphene in
particular, other exotic manifestations of the physics of
zero-modes become possible due to their  rich internal structure,
consisting of spin, valley, sublattice, and the Nambu degrees of
freedom\cite{HCM, IgorInsulating,Igor2010,ghaemi}. It is therefore
of importance to understand the general conditions for the
appearance of the zero-energy states. Quite generally, we find
that the zero-modes in the superconducting vortex may exist only
for weak enough Zeeman coupling to the magnetic field. This is in
accord with our previous study limited to the the vortex in the
s-wave superconducting order parameter\cite{HerbutLu}.

This paper has the following organization. We first
investigate the spectrum associated with the spatially homogeneous
Hamiltonian without vortex for the singlet $s$-wave, triplet
$p$-wave and mixed-symmetry cases. The conditions for opening of a
gap in all three cases are discussed. We then  proceed to
the equations for the zero-mode associated with different pairing
symmetries, with the Zeeman term and the chemical potential
included. In the sec. IV, the effects of the Zeeman field and the
relation of the zero-mode with the antilinear operator that
provides the reflection symmetry of the spectrum\cite{HerbutLu} in
all symmetry cases are discussed. Finally, a brief summary is
given.

\section{Translation-invariant massive Dirac Hamiltonian}

\begin{figure}
\input{epsf}
\includegraphics[scale=0.5]{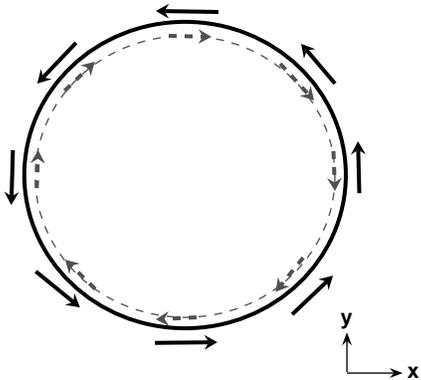}
\caption{A single two-dimensional Fermi surface associated with
the Dirac Hamiltonian,
$H_0(\vec{k})=v_F(\hat{\rm{z}}\cdot{\vec{k}\times{\vec{\sigma}}})-\mu$,
on the metallic surface of a TI. $\hat{\rm{z}}$, the out-of-plane
direction, is the spin quantization axis. The dark (dashes) arrows
indicate the directions of the spin associated with the momentum
state $\vec{k}$ on the Fermi surface of positive (negative)
helicity when $\mu$ is positive (negative).}\label{SingleFS}
\end{figure}

In this section, we consider the Dirac electrons on the metallic
surface of a TI and their pairing due to the proximity to a
superconducting material. The metallic surface of a TI has a
single Fermi surface as shown in Fig. \ref{SingleFS}, and the
spin-dependent spectrum can be described by a simple 2$\times$2
Hamiltonian,
$H_0(\vec{k})=v_F\hat{\rm{z}}\cdot\vec{k}\times\vec{\sigma}-\mu$,
where $\mu$ is the chemical potential and $v_F$ is the Fermi
velocity. The quantization axis for spin is parallel with the
normal of the surface $\hat{\rm{z}}$. As usual, the identity
matrix and the three Pauli matrices, $\{I_2,\vec{\sigma}\}$,
define a basis in the two-dimensional matrix space. One should
note that $H_0(\vec{k})$ is invariant under the point group
C$_{4{v}}$ in which the inversion symmetry is
absent\cite{YipGarg}. At the mean-field level, the general pairing
potential $V_p$ that respects the absence of inversion symmetry
and takes different coupling channels (spin singlet and triplet)
into account needs to be be expressed as a matrix.\cite{Gorkov} To
this purpose we introduce the general mass matrix
$M(\vec{k})=\Delta{I_2}+v_{\Delta}\vec{d}(\vec{k})\cdot\vec{\sigma}$,
so that the pairing potential becomes
$V_p(\vec{k})=iM(\vec{k})\sigma_y$. Here $\Delta$ stands for the
singlet pairing gap while the velocity $v_{\Delta}$ is associated
with the triplet pairing.

In the 4-dimensional spin-Nambu space, the BdG Hamiltonian
$\hat{H}=\sum_{\vec{k}}\Psi^{\dag}_{\vec{k}}\hat{H}(\vec{k})\Psi_{\vec{k}}$, with
$\Psi_{\vec{k}}=\{a_{k,\uparrow},a_{k,\downarrow},a^{\dag}_{-k,\downarrow},-a^{\dag}_{-k,\uparrow}\}^T$,
then assumes the form,

\be
    \hat{H}(\vec{k})=\left(\begin{array}{cccc}
    &H_0(\vec{k})&M(\vec{k})\\
    &M^{\dag}(\vec{k})&-H_0(\vec{k})
    \end{array}\right)\:.\label{fullHamiltonian}
\ee For later convenience, we note that $\hat{H}(\vec{k})$ can
also be written in terms of the cross-product representation,
namely
$\hat{H}(\vec{k})=(v_F\vec{n}(\vec{k})\cdot\vec{\sigma}-\mu)\tau_3+M(\vec{k})\tau_{+}
+M^{\dag}(\vec{k})\tau_{-}$, where $\sigma_i$'s act on the spin
and  $\tau_i$'s on the Nambu indices. The vector
$\vec{n}(\vec{k})=(-k_y,k_x)$ lies on the x-y plane, as in Fig.
\ref{SingleFS}. The chemical potential $\mu$ enters along with the
number operator $\rm{N}$, which can be identified as $\tau_3$ in
this representation. As we discuss shortly, while they  of course
never commute with the number operator $\tau_3$, the mass terms
may commute with the diagonal, kinetic energy terms. These cases
will lead to a peculiar spectrum that is not possible in standard
superconductors which have the quadratic energy-momentum
dispersion in its normal state.

First, let us discuss different pairing symmetries for the mass
term, in the translation-invariant situation without the vortex,
and the resulting  spectrum $E(\vec{k})$.

The simplest case corresponds to the singlet s-wave pairing in
which $\Delta$ is constant and $v_{\Delta}$=0. The spectrum may be
obtained most easily by squaring the Hamiltonian and using the
anticommuting properties of the Pauli matrices; this gives
$[\hat{H}(\vec{k})]^2=(v_F\vec{n}(\vec{k})\cdot\vec{\sigma}-\mu)^2+M^2$,
and therefore
$E(\vec{k})=\pm\sqrt{(v_Fk\pm\mu)^2+\Delta^2}$.\cite{FuKane,Santos}
Next we consider the triplet pairing, with the vector $\vec{d}$
coinciding with $\vec{n}$, in which the mass term
$M=v_{\Delta}\vec{n}\cdot\vec{\sigma}$ still anticommutes with the
kinetic energy. The same trick can be applied again, to obtain the
spectrum,

\be
    E(\vec{k})=\pm\sqrt{(v_Fk\mp\mu)^2+(v_{\Delta}k)^2}\:,\label{TripletSpec}
\ee which is gapped as long as $\mu$ is finite. At $\mu=0$, the
spectrum is gapless, and in fact reduces to the Dirac cone, with
an increased velocity. A somewhat unusual situation arises when
$\vec{d}=(k_x,k_y)$, for which the mass term {\it commutes} with
the kinetic energy: when $\vec{n}\bot\vec{d}$,
$[M\tau_{\pm},\vec{n}\cdot\vec{\sigma}\tau_3]$=0. It can be shown
that the spectrum is then given by
$E(\vec{k})=\pm{v_F}k\pm\sqrt{\mu^2+(v_{\Delta}k)^2}$. In this
case the spectrum is still gapless,  and it is the chemical
potential that becomes modified and momentum-dependent, loosely
speaking\cite{Linder}.

In the absence of the inversion symmetry  the general ``mixed"
mass term, with both $\Delta$ and $v_{\Delta}$ finite, is also
allowed. Let us assume  the triplet component to be given by
$\vec{d}=\vec{n}$. By squaring the Hamiltonian, one can obtain,
$[\hat{H}(\vec{k})]^2=(v_F\vec{n}\cdot\vec{\sigma}-\mu)^2+(\Delta_s+v_{\Delta}\vec{n}\cdot\vec{\sigma})^2$,
and consequently

\be
    E(\vec{k})=\pm\sqrt{(v_Fk\mp\mu)^2+(\Delta\pm{v_{\Delta}}k)^2}\:,\label{MixedSpec}
\ee where a gapped spectrum exists only for the positive helicity
branch when $\Delta=v_{\Delta}k_F=\Delta_{+}/2$, with the Fermi
wavevector $k_F=\mu/v_F$.

\section{Equations for zero-energy quasi-particle state}

In this section we shall consider the inhomogeneity in the mass
term $M$ generated by an isolated vortex in which the magnetic
field applied along $\hat{\rm{z}}$ is threading through its core.
The superconducting phase changes by $2\pi{Q}$ on circling once
around the core. Here $Q$ labels the winding number of the vortex.
In the context of superconductivity, the mass term here is in
general a function of both the center-of-mass position $\vec{r}$
and relative momentum $\vec{k}$ of the Cooper pair, namely,
$M=M(\vec{r},\vec{k})$. More explicitly, in the polar coordinate
$\vec{r}=(r\cos\phi,r\sin\phi)$, the vortex enters the mass via
$M(\vec{r},\vec{k})=e^{{\it{i}}Q\phi}M(\vec{k})$. In the following
subsections, we shall employ the BdG equation to investigate the
zero-energy bound state in the vortex core for various pairing
symmetries. In particular, we also consider the effects of an
external Zeeman field
$H_{\rm{Z}}=h\sum_{\vec{k}}\Psi^{\dag}_{\vec{k}}(\sigma_3\tau_0)\Psi_{\vec{k}}$,
in which $\tau_0$ stands for the identity matrix acting on the
Nambu space.

\subsection{Singlet $s$-wave pairing}\label{singlet}

The mass generated by the vortex of singlet
$s$-wave pairing potential with the  winding number of unity can be
written as $M=e^{i\phi}|\Delta(r)|$, in which the magnitude
increases from zero at the core and reaches a positive value of
$\Delta_{\infty}$ when the distance $r$ is sufficiently large than
the superconducting coherence length. In terms of the Dirac fermion
$\Psi$ we defined, the quasi-particle states which diagonalize the BdG
Hamiltonian and give exactly zero-energy have the corresponding
coefficients ${\bf{u}}=\{u_1(\vec{r}),u_2(\vec{r})\}$ for the electron
sector and the corresponding components ${\bf{v}}$ for the hole sector
satisfy the following equations,

\bea
    \left[-\mu+v_F\hat{\rm{z}}\cdot(\vec{k}-e\vec{A})\times{\vec{\sigma}}
    +h\sigma_z\right]{\bf{u}}
    +e^{i\phi}|\Delta(r)|{\bf{v}}=0\:,\\
    e^{-i\phi}|\Delta(r)|{\bf{u}}+\left[\mu-v_F\hat{\rm{z}}\cdot(\vec{k}+e\vec{A})\times{\vec{\sigma}}
    +h\sigma_z\right]{\bf{v}}=0\:,
\eea where $\hbar$=$c$=1 is taken for simplicity. The momentum
operators, $k_i=-i\partial_{i}$, are coupled to the gauge field
$\vec{A}=(-y,x)B(r)/2$, where $B$ varies on a
length scale of penetration length $l_p$ inside a superconductor.
Using the identity
$(\partial_x\pm{i}\partial_y)$=$e^{\pm{i}\phi}(\partial_r\pm (i/r)\partial_{\phi})$,
and the angular decomposition,
$[{\bf{u}},{\bf{v}}]^T$=$[u_1(r),e^{i\phi}u_2(r),e^{-i\phi}v_1(r),v_2(r)]^T$,
one can decouple the set of four first-order differential
equations into two blocks, namely,

\bea
    \left[v_F(\partial_r+\frac{1}{r}-\frac{r}{2}B)-|\Delta|\right]f_a
    +(\mu-h)f_b=0\:,\label{Forb1}\\
    \left[v_F(\partial_r+\frac{r}{2}B)-|\Delta|\right]f_b
    -(\mu+h)f_a=0\:,\label{Forb2}\\
    \left[v_F(\partial_r+\frac{1}{r}-\frac{r}{2}B)+|\Delta|\right]g_a
    +(\mu-h)g_b=0\:,\label{decay1}\\
    \left[v_F(\partial_r+\frac{r}{2}B)+|\Delta|\right]g_b
    -(\mu+h)g_a=0\:,\label{decay2}
\eea where the definitions $u_2\pm{v_1}$=$f_a$($g_a$) and
$u_1\mp{v_2}$=$f_b$($g_b$) are made. The explicit dependence of
magnetic field $B$ and pairing potential $\Delta$ on the radial
coordinate $r$ is not shown to simplify the notation. In fact, it
will be seen in the following asymptotic analysis that the effects
of coupling to the gauge field on the zero-mode are negligible.


Far from the core the gap function
$\Delta(r)\rightarrow\Delta_{\infty}$ and moreover, only the
derivative terms are relevant in the kinetic energy. The terms
related to $B(r)$ can be dropped since $B(r)$ decay faster than
$1/r$ in the case of a isolated vortex\cite{deGennesB}. Therefore,
by defining the new variable $x\equiv\sqrt{|\mu^2-h^2|}r/v_F$,
Eqs. (\ref{Forb1}) and (\ref{Forb2}) are reduced to the following:

\be
    (\frac{d^2}{dx^2}\pm1)\left[e^{-\frac{\Delta_{\infty}}{\sqrt{|\mu^2-h^2|}}x}f_a(x)\right]=0\:,
\ee where the plus(negative) signs are indicative of that $|\mu|$
is larger(smaller) than $|h|$, respectively. In both cases, as
long as $h^2<\mu^2+\Delta^2_{\infty}$, the two solutions for $f_a$
are both exploding at infinity, which, from Eq.(\ref{Forb1}), is
also true for $f_b$. Thus one concludes that $u_1=v_2$ and
$u_2=-v_1$, to ensure the normalizability. The remaining
components can be determined then from Eq. (\ref{decay1}) and
(\ref{decay2}), which can be reduced to,

\be
    (\frac{d^2}{dx^2}\pm1)\left[e^{\frac{\Delta_{\infty}}{\sqrt{|\mu^2-h^2|}}x}g_a(x)\right]=0\:,
\ee where two decaying solutions are available under the same
condition $h^2<\mu^2+\Delta^2_{\infty}$.

Near the core where the argument $x\ll1$, it is legitimate to keep
only the singular and the derivative terms. This is because
$\Delta(r)$ is vanishingly small there, and the inherent
logarithmic divergence from $B(r)$\cite{deGennesB} is suppressed
in the vector potential. Eqs. (\ref{decay1}) and (\ref{decay2})
can also be written as,

\be
    (\frac{d^2}{dx^2}+\frac{1}{x}\frac{d}{dx}-\frac{1}{x^2}\pm{1})g_a(x)=0\:,\label{CoreEq}
\ee where the plus sign still means that $|\mu|$ is larger than
$|h|$. Now one can see that $u_2(x)$=$J_1(x)$, the first-order
Bessel function, for small argument. Note that the other solution,
$N_1(x)$, is forbidden because of its singularity of $1/x$ for
small argument. For $|\mu|<|h|$, $u_2(x)$=$I_1(x)$, the modified
Bessel function which is regular at small $x$. The other component
$u_1(x)$ is not independent and is given by,

\be
    u_1(x)=-\frac{\sqrt{|\mu^2-h^2|}}{\mu-h}(\frac{d}{dx}+\frac{1}{x})u_2(x)
    \:,
\ee for small argument $x$. If we think of the pair potential
$\Delta(r)$ as a step function which vanishes within some certain
range and is a constant outside, the obtained solution from the
second-order differential equation for $x\ll1$ can be smoothly
joined with the solution for $x\gg1$ at the boundary by matching
the values of the function and its first derivative. This can be
accomplished when $h^2<\mu^2+\Delta^2_{\infty}$ since, given one
function on the left-hand side of the boundary, there are two
available decaying solutions on the right-hand side. For an
arbitrary $\Delta(r)$ the solutions can be obtained by
adiabatic deformation of the step-like pair potential  that leaves the
asymptotic behaviors intact.

We conclude that, as long as $h^2<\mu^2+\Delta^2_{\infty}$, there
exists one zero-energy solution for the $s$-wave pairing vortex of
winding number one, with the following radial wave function,

\be
    \left(\begin{array}{cccc}
    u_1(r)\\
    u_2(r)\\
    -u_2(r)\\
    u_1(r)
    \end{array}\right)\:,\label{S_wave_function}
\ee where the electron components $u_1$ and $u_2$ for small argument are given by,

\be
    \left(\begin{array}{cccc}
    u_1(x)\\
    u_2(x)
    \end{array}\right)=
    \left(\begin{array}{cccc}
    \frac{\sqrt{|\mu^2-h^2|}}{\mu-h}J_0(x)\\
    -J_1(x)
    \end{array}\right)\:,\label{s_wf_1}
\ee when $h^2<\mu^2$. As for $\mu^2<h^2<\mu^2+\Delta^2_{\infty}$,
$J_0(x)$ and $J_1(x)$ are simply replaced with the corresponding
modified Bessel functions $I_0(x)$ and $I_1(x)$, respectively. For
a large argument,

\be
    \left(\begin{array}{cccc}
    u_1(x)\\
    u_2(x)
    \end{array}\right)=
    \left(\begin{array}{cccc}
    \frac{\sqrt{|\mu^2-h^2|}}{\mu-h}(A\cos{x}-B\sin{x})\\
    -(A\sin{x}+B\cos{x})
    \end{array}\right)e^{-\frac{\Delta_{\infty}}{\sqrt{|\mu^2-h^2|}}x}\:,\label{s_wf_2}
\ee which is valid only for $h^2<\mu^2$. For
$\mu^2<h^2<\mu^2+\Delta^2_{\infty}$, the trigonometric  functions
are replaced with the corresponding hyperbolic ones. The
coefficients $A$ and $B$ are to match $u_2(x)$ at the boundary and
an additional overall coefficient for normalization is not shown
explicitly.

A special case of above is when $\mu=h=0$ with the solution given
by,

\be
    \left(\begin{array}{cccc}
    u_1(r)\\
    u_2(r)
    \end{array}\right)=
    \left(\begin{array}{cccc}
    1\\
    0
    \end{array}\right)e^{-\frac{1}{v_F}\int^r|\Delta(r')|dr'}\:,\label{s_wf_FuKane}
\ee which is exact for all range of $r$. Comparing with Eqs. \
(\ref{s_wf_1}) and (\ref{s_wf_2}), one should note that the
ambiguity arising at the zero
chemical potential is removed by the fact that $J_1$ is
vanishing for zero argument. Besides, the zero-mode in presence of
a finite chemical potential appears with both spin projections in
contrast to the zero-mode in Eq.\ (\ref{s_wf_FuKane}) where only
the up-spin components show up\cite{FuKane}. For the case with an
anti-vortex, $M\sim|\Delta(r)|e^{-i\phi}$, different angular
decomposition,
$[{\bf{u}},{\bf{v}}]^T$=$[e^{-i\phi}u_1(r),u_2(r),v_1(r),e^{i\phi}v_2(r)]^T$,
has to be employed, and, consequently, $u_1\sim{J_1}$ and
$u_2\sim{J_0}$ near the core. Moreover, in the special case with
vanishing $\mu$ and $h$, the Majorana zero-mode have similar form
as in Eq.\ (\ref{s_wf_FuKane}) but with down-spin components only.

\subsection{Triplet $p$-wave pairing}\label{triplet}

Now we consider the triplet order parameter specified by
$\vec{d}=(-k_y,k_x)$, which is parallel to the spin direction
$\vec{n}$ in the kinetic energy. With this Rashba type $p$-wave
pairing, an isotropic superconducting gap is generated for nonzero
$\mu$ according to Eq.\ (\ref{TripletSpec}) in contrast to the
case with $\vec d\perp\vec n$. The quasi-particle states with zero
energy in the presence of vortex in the triplet order parameter
with winding number of unity have to satisfy the following
equations,



\bea
    v_F(\partial_r+\frac{1}{r})f_a+(\mu-h-v_{\Delta}\partial_r)f_b=0\label{f1}\:,\\
    v_F\partial_rf_b-\left[\mu+h-v_{\Delta}(\partial_r+\frac{1}{r})\right]f_a=0\label{f2}\:,\\
    v_F(\partial_r+\frac{1}{r})g_a+(\mu-h+v_{\Delta}\partial_r)g_b=0\label{g1}\:,\\
    v_F\partial_rg_b-\left[\mu+h+v_{\Delta}(\partial_r+\frac{1}{r})\right]g_a=0\label{g2}\:,
\eea where identical angular dependence has been employed, and
$f_{a(b)}$ and $g_{a(b)}$ have the same definitions as before.
$v_{\Delta}(r)$ is also a radial function. When $\mu=h=0$, it can
readily be deduced that both $f_a$ and $g_a$ must be identically zero; otherwise
the operator $(\partial_r+1/r)$ will lead to the divergence of
$\sim 1/r$ in the solution near the origin. Consequently, the
zero-energy state must satisfy $\partial_rf_b=\partial_rg_b=0$ for
all $r$, which leads to the conclusion that the zero-energy bound
state does not exist when  $\mu=0$. For general $\mu$ and $h$, on the other hand,
we may proceed in the same manner as in previous subsection. Near the vortex
core the analysis is identical
 since the gap velocity $v_{\Delta}$ is vanishingly small
there. Far from the core, it is useful to  rewrite Eq. (\ref{f1})
and (\ref{f2}) in terms of $f_a\pm{i}f_b$, as,


\bea
    \left[(v_F+iv_{\Delta})\partial_r-i\mu\right](f_a+if_b)-ih(f-if_b)=0\:,\label{faa}\\
    \left[(v_F-iv_{\Delta})\partial_r+i\mu\right](f_a-if_b)+ih(f+if_b)=0\:,\label{fbb}
\eea which also can be recast as  a second-order differential
equation. For $\mu^2-h^2>0$, the desired equation is
$(d^2/dy^2-2\lambda{d}/dy+1)(f_a+if_b)=0$, where the variable
$y=\sqrt{|\mu^2-h^2|/(v_F^2+v_{\Delta}^2)}r$ and the constant
$\lambda=\mu{v_{\Delta}}/\sqrt{(v_F^2+v_{\Delta}^2)|\mu^2-h^2|}$.
The solution then has a plane wave form, $e^{iky}$, with
$k=-i\lambda\pm\sqrt{1-\lambda^2}$. Without loss of generality
$v_{\Delta}$ may be assumed to be positive. We then first consider $\mu$
to be positive: there are then two exploding solutions for $f_a+if_b$ at
infinity when $0<\lambda<1$, or equivalently
$\mu^2>(1+v^2_{\Delta}/v^2_F)h^2$. For the case of $\lambda>1$, or
$h^2$ is more close to $\mu^2$, the two solutions are still
exploding but with $k=-i(\lambda\mp\sqrt{\lambda^2-1})$. On the
other hand, for $\mu^2<h^2$, the differential equation becomes
$(d^2/dy^2-2\lambda{d}/dy-1)(f_a+if_b)=0$, with the asymptotic
solution of wave vector $k=-i(\lambda\pm\sqrt{\lambda^2+1})$. In this case
one solution is always decaying, while the other is exploding. Thus we
conclude that $f_a=f_b=0$ as long as $\mu^2>h^2$,  and we take
$u_1=v_2$ and $u_2=-v_1$. The remaining task then
is to obtain the $u_i$'s from the equations for $g_a$ and $g_b$,
which are identical to Eqs. (\ref{faa}) and (\ref{fbb}), except that
$v_{\Delta}\rightarrow-v_{\Delta}$ and hence two decaying
solutions at infinity are available as long as $\mu^2>h^2$.

For $\mu$ negative, the above conclusion has to be reversed, so
that $g_a=g_b=0$ when $\mu^2>h^2$, and therefore $u_1=-v_2$ and
$u_2=v_1$. The rest of analysis is the same as with positive $\mu$.
The complete wave function for zero-energy state has
the radial part summarized as following:

\be
    \left(\begin{array}{cccc}
    u_1(r)\\
    u_2(r)\\
    -{\rm{sgn}}(\mu)u_2(r)\\
    {\rm{sgn}}(\mu)u_1(r)
    \end{array}\right)\:,\label{p_wf_1}
\ee and the asymptotic behaviors are given by

\be
    \left(\begin{array}{cccc}
    u_1(y)\\
    u_2(y)
    \end{array}\right)=
    \left(\begin{array}{cccc}
    (A-B)\sin{\sqrt{1-\lambda^2}y}\\
    (A+B)\cos{\sqrt{1-\lambda^2}y}
    \end{array}\right)e^{-|\lambda|y}\:,\label{p_wf_2}
\ee which has an oscillation on top of a decaying form when
$0<\lambda<1$. It will be monotonically decaying at infinity when
$\lambda>1$, and the trigonometric functions need to be  replaced with the
corresponding hypergeometric ones. The coefficients are determined by matching
the function $u_2+iu_1$ to the solution  near the core.

\subsection{Mixed $s$-wave and $p$-wave pairing}\label{mixed}

Finally, we consider the combination of the problems in the previous two
subsections. In absence of the inversion symmetry the kinetic
energy in the normal state can have its chiral form  with the
pairing potential as the mixture in two order parameters of different parities.
We will thus assume it to be  a mixture of the singlet and
triplet components,  with the latter specified by
$\vec{d}=(-k_y,k_x)$. For simplicity, we also neglect the Zeeman effect.
Far from the vortex core, the corresponding equations
for $f$'s are, similarly as before,

\bea
    (v_F\partial_r-\Delta_{\infty})f_a+(\alpha|\mu|-v_{\Delta}\partial_r)f_b=0\:,\\
    (v_F\partial_r-\Delta_{\infty})f_b-(\alpha|\mu|-v_{\Delta}\partial_r)f_a=0\:,
\eea in which $\alpha$ specifies the sign of chemical
potential,  and we take $\alpha=1$ first. The above equations are
identical to those for $g_a$ and $g_b$, except that
$\Delta_{\infty}\rightarrow-\Delta_{\infty}$ and
$v_{\Delta}\rightarrow-v_{\Delta}$. It follows that at
infinity both $f_a\pm{if_b}$ are in the form of
$e^{(\kappa\pm{iq}){r}}$ with
$\kappa=(\alpha|\mu|{v_{\Delta}+\Delta_{\infty}v_F})/(v^2_{\Delta}+v^2_F)$
positive,  and
$q=(\alpha|\mu|{v_F-\Delta_{\infty}v_{\Delta}})/(v^2_{\Delta}+v^2_F)$.
It is useful to define the parameters

\be
    \Delta_{\pm}=\Delta_{\infty}\pm{v}_{\Delta}{k}_F\:,\label{DefDeltaPM}
\ee where $v_Fk_F=|\mu|$ is the Fermi momentum for nonzero
chemical potential and the renormalized Fermi velocity
$\bar{v}_F=v_F(1+v^2_{\Delta}/v^2_F)$. The exploding
solutions can then be written as $\sim e^{\Delta_{+}r/{\bar v}_F}$, which is
similar to the case of $s$-wave pairing. Of course, the exploding solutions
are forbidden, and therefore $u_1=v_2$ and $u_2=-v_1$. It can be
seen that the set of equations for $g_a$ and $g_b$ will lead to
two asymptotic decaying solutions in the form
$e^{-\Delta_{+}r/{\bar v}_F}e^{\pm{iqr}}$, which in turn
determines the zero-mode wavefunctions to be,

\be
    \left(\begin{array}{cccc}
    u_1(r)\\
    u_2(r)
    \end{array}\right)=
    \left(\begin{array}{cccc}
    Ae^{i(qr+\frac{\pi}{2})}-Be^{-i(qr-\frac{\pi}{2})}\\
    Ae^{iqr}+Be^{-iqr}
    \end{array}\right)e^{-\Delta_{+}{r}/\bar{v}_F}\:,\label{mixed_wf}
\ee with the coefficients determined by the similar boundary
conditions for matching $u_2$ to its piece near the core.

For a negative chemical potential, by similar
reasoning, the set of equations associated with $f$'s gives two
asymptotic solutions as $e^{\Delta_{-}r/{\bar v}_F}e^{\pm{iqr}}$,
which will be forbidden if $\Delta_{-}>0$. Then, the zero-mode is
determined from the set of equations associated with the $g$'s,
which give similar solution as in Eq.\ (\ref{mixed_wf}), except
that $\Delta_{+}\rightarrow\Delta_{-}$. However, the zero-mode
will have $u_1=-v_2$ and $u_2=v_1$ in contrast to the previous
case when $\Delta_{-}<0$, in which the solutions from the $f$'s
were decaying and allowed, while those from $g$'s were forbidden.
The wavefunction for $\mu<0$ therefore has a general form,

\be
    \left(\begin{array}{cccc}
    u_1(r)\\
    u_2(r)\\
    -{\rm{sgn}}(\Delta_{-})u_2(r)\\
    {\rm{sgn}}(\Delta_{-})u_1(r)
    \end{array}\right)\:,\label{mix_wf_1}
\ee where one can see that a vanishing $\Delta_{-}$ will remove
the zero-mode. The corresponding $u_1$ and $u_2$ are determined in
a similar manner.

\section{Discussions and Conclusions}

In the preceding sections, the existence of zero-energy bound
states in the presence of various symmetries of the mass-gap is
investigated in the context of superconducting Dirac fermions on
the surface of a TI. It may also be  useful to write the BdG
Hamiltonian in presence  of the vortex in terms of the
4-dimensional Dirac $\Gamma$-matrices. The $s$-wave vortex, for
example, is described by the Hamiltonian
$\hat{H}=\Gamma_1p_1+\Gamma_2p_2+|\Delta(r)|(\Gamma_3\cos\phi+\Gamma_5\sin\phi)+ih\Gamma_1\Gamma_2-i\mu\Gamma_3\Gamma_5$.
The last two terms are the Zeeman field and the chemical
potential, respectively. Choosing the representation for the
matrices as $\Gamma_1=\sigma_z\otimes\sigma_y$,
$\Gamma_2=-\sigma_z\otimes\sigma_x$,
$\Gamma_3=\sigma_x\otimes{I_2}$, and
$\Gamma_5=-\sigma_y\otimes{I_2}$ will then yield the original
differential equations. The four matrices, together with $\Gamma_0
= \sigma_z\otimes\sigma_z$, are all Hermitian, and anticommuting
among themselves, $\{\Gamma_i,\Gamma_j\}=2\delta_{ij}$ for
$i,j=0,1,2,3,5$.  It is easy to see that in the absence of the
last two terms the matrix $\Gamma_0$ anticommutes with the
Hamiltonian, and thus ensures that the spectrum has the reflection
symmetry around zero. In this limit there is a total of $|n|$
zero-modes for the vortex of vorticity $n$\cite{Jackiw1,IndexW}.
It can be shown from the general properties of the Clifford
algebra satisfied by the Dirac matrices that there also exists one
antilinear operator that anticommutes with $\hat{H}$ even in the
presence of the last two, $\sim \Gamma_1\Gamma_2$ and $\sim
\Gamma_3\Gamma_5$ terms\cite{footnote}. In the representation
defined above, in which $\Gamma_1$ and $\Gamma_5$ are the only two
matrices among five which are imaginary, the antilinear operator
in question is $A=i\Gamma_1\Gamma_3K$, where $K$ denotes the
operation of complex conjugation. Explicitly,



\be
    A=\left(\begin{array}{cccc}
    0&i\sigma_y\\
    -i\sigma_y&0
    \end{array}\right)K\:.
\ee We recognize that the zero-mode $\Psi_0$ in Eq.\
(\ref{S_wave_function}) is its eigenvector with eigenvalue $1$:
$A\Psi_0=\Psi_0$. Here, it is worthwhile noting that $\Gamma_0$
and $A$ serve as the sublattice (chiral) and particle-hole
symmetry operators, respectively. In the presence of finite
chemical potential and/or Zeeman field, only the particle-hole
symmetry is present and the JRD Hamiltonian belongs to the class D
in the symmetry classifications.\cite{Altland} In the case of
vanishing chemical potential and zero magnetic field, the chiral
symmetry is restored and the JRD Hamiltonian belongs to the class
DIII.





\begin{figure}
\input{epsf}
\includegraphics[scale=0.5]{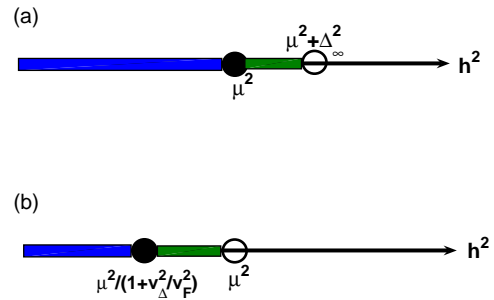}
\caption{(Color online) The phases of zero-mode for $h^2$ versus
$\mu^2$ in the (a) $s$-wave pairing and (b) $p$-wave pairing
cases. The left-handed sides with respect to the empty circles
denote the existence of the zero-mode, while the right-handed
sides, including the empty circle, indicate that the zero-mode is
not normalizable and hence does not exist. The blue regions stand
for the fact that the wavefunction has, in addition to the
decaying, an asymptotic oscillation part, which is gone when
entering the green regions.}\label{PurePhase}
\end{figure}

The effects of the Zeeman field on the zero-mode may be summarized
as in Fig. \ref{PurePhase}(a). For
$|h|>\sqrt{\mu^2+\Delta^2_{\infty}}$, specified by the empty
circle, the zero-mode is not normalizable, and hence disappears
from the spectrum. The full circle, on the other hand,  denotes
the end of oscillations in  the zero-mode in Eq.\ (\ref{s_wf_1})
and (\ref{s_wf_2}). Lastly, the presence of a single Fermi surface
is the key difference in comparison with the same problem of the
vortex in a noncentrosymmetric superconductor,  in which the
spin-orbit coupling is just the chiral kinetic energy, while the
ordinary kinetic energy enter the Hamiltonian as
$i(k^2/2m-\mu)\Gamma_3\Gamma_5$. It is easy to see that while the
operator $A$ still anticommutes with the Hamiltonian, no zero-mode
exists in the case for the $s$-wave symmetry of the order
parameter\cite{Lu2}.


In the triplet case, the corresponding BdG Hamiltonian has a
momentum-dependent mass term. In the Dirac notation, the
Hamiltonian is now
$\hat{H}=H_D+iMH_D+ih\Gamma_1\Gamma_2-i\mu\Gamma_3\Gamma_5$, where
$H_D$ and $M$ are of the same form as, respectively, the kinetic
energy and $s$-wave mass in the previous paragraph. It can be
shown that the operator $A$ still facilitates the reflection
symmetry of the energy spectrum, due to the  factor of $i$ in the
order-parameter term, which on the other hand is necessary to make
it Hermitian. We showed here that the zero-mode does not exist
when $\mu=0$, which can also be understood in terms of the
symmetry of the zero-mode $\Psi_0$ in Eq.\ (\ref{p_wf_1}) under
$A$: for positive and negative chemical potential $\mu$, we find
that $A\Psi_0=\pm\Psi_0$, respectively. In fact, the disappearance
of the zero-mode at $\mu=0$ suggests a topological
transition\cite{DasSarma}, because the spectrum in Eq.\
(\ref{TripletSpec}) in the translationally invariant case becomes
gapless right at $\mu=0$. In Fig. \ref{PurePhase}(b) we summarize
the effects of Zeeman field on the zero-mode. For $h^2>\mu^2$,
there is no zero-mode,  as again denoted  by the empty circle.
Similarly, the full circle at $h^2=\mu^2/(1+v^2_{\Delta}/v^2_F)$,
represents the disappearance of oscillations under the overall
exponential decaying in the zero-mode.


In a real type-II superconductor, the Zeeman field $h$ is not
uniform and is proportional to the distribution of magnetic field
$B(r)$, a function riding on the scale of penetration length
$l_p$. The obtained zero-mode $\Psi_0$ possesses two length scales
$k_F^{-1}=v_F/\sqrt{|\mu^2-h^2|}$ and $\xi=v_F/\Delta$, which are
assumed to be much smaller than the penetration length in the
present problem. Therefore, it is valid to regard $h(r)$ as a
\emph{quasi}-constant throughout the space. In fact, the
distribution $B(r)=\frac{\phi_0}{2\pi{l}_p^2}K_0(r/\lambda)$ for
an isolated vortex, where $\phi_0$ is the magnetic flux. Hence, it
could happen that near the core $|h|>|\mu|$ as a result of
$|B|\sim\ln\frac{{l}_p}{r}$ when $r\ll{l}_p$\cite{deGennesB}.
Correspondingly, the zero-mode will behave monotonically near the
core, which is just the same as the the modified Bessel function
with a length scale of $v_F/|h|$. Nevertheless, the zero-mode will
still be sustainable with the asymptotic $h(r\rightarrow\infty)$
within the range in Fig.\ref{PurePhase}.

\begin{figure}
\input{epsf}
\includegraphics[scale=0.5]{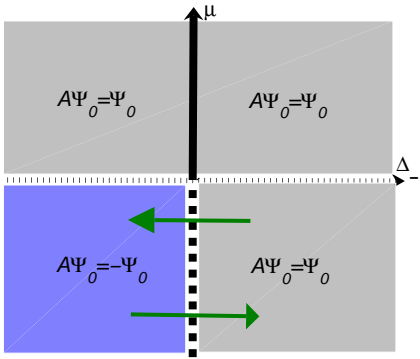}
\caption{(Color online) Different phases for the zero-mode in the
mixed pairing symmetry, as a function of $\mu$ and the parameter
$\Delta_{-}$, assuming $\Delta_{+}$ is positive. The dotted axis
is to exclude the special cases with $\mu=0$. The grey and blue
regions denote different eigenvalues of $A$ for the zero-mode
$\Psi_0$ as specified in text. At the dashed line $\mu<0$,
$\Delta_- =0$, in between the two phases where $\Delta_-$ changes
sign (as denoted by the green arrows) the localized zero-mode does
not exist.}\label{MixedPhase}
\end{figure}

The mixed $s$-wave and $p$-wave symmetry is of relevance when it
comes to the 2D surface considered here. In this case, the
corresponding mass term is $M+iMH_D$, which still anticommutes
with $A$ and the principles outlined above can be applied. The
amplitudes of the respective pairings are determined by their
asymptotic values $\Delta_{\infty}$ and $v_{\Delta}$. First, note
that the Fermi surfaces for $\mu>0$ and $\mu<0$ have opposite
helicities, and the corresponding zero-mode wave functions in Eq.\
(\ref{mixed_wf}) have the decaying length as ${\bar
v}_F/\Delta_{+}$ and ${\bar v}_F/|\Delta_{-}|$, respectively. As
shown in Fig. \ref{MixedPhase}, the eigenvalue $\lambda$ of the
zero-mode, $A\Psi_0=\lambda\Psi_0$, may be different for opposite
signs of the product $\Delta_{+}\Delta_{-}$. Assuming $\mu$ is
nonzero and $\Delta_{+}$ is positive, $\lambda$ remains the same
when one changes $\mu=|\mu|$ to $\mu=-|\mu|$ if $\Delta_{-}>0$,
which is similar to the $s$-wave case. However, $\lambda$ changes
sign if $\Delta_{-} < 0$, and this resembles the situation in the
$p$-wave case. Besides, the zero-mode stops being normalizable
when $\mu<0$ and the parameters $\Delta_{\infty}$ and
$k_Fv_{\Delta}$ have $\Delta_{-}$ vanishing, which corresponds to
a gapless condition in Eq.\ (\ref{MixedSpec}). This is specified
by the dashed line in Fig. \ref{MixedPhase}. The meaning of the
parameters $\Delta_{\pm}$ is clear when one considers the helicity
basis, $\{a^{\dag}_{\rm{k}+},a^{\dag}_{\rm{k}-}\}$, which can be
transformed back to the ordinary basis,
$\{a^{\dag}_{\rm{k}\uparrow},a^{\dag}_{\rm{k}\downarrow}\}$, with
the following,

\be
    \left(\begin{array}{cccc}
    a^{\dag}_{\rm{k}+}\\
    a^{\dag}_{\rm{k}-}
    \end{array}\right)=\frac{1}{\sqrt{2}}
    \left(\begin{array}{cccc}
    1 &ie^{i\phi_{\bf k}}\\
    ie^{-i\phi_{\bf k}}& 1
    \end{array}\right)
    \left(\begin{array}{cccc}
    a^{\dag}_{\rm{k}\uparrow}\\
    a^{\dag}_{\rm{k}\downarrow}
    \end{array}\right)\:,
\ee where the angle $\phi_{\bf k}$ denotes the direction
associated with the momentum on the
plane.\cite{YipTopology,Santos} Therefore, the time-reversal
invariant pairing Hamiltonian can also be written
as,\cite{YipTopology}

\be
    H_{\Delta}=\frac{1}{2}\sum_{\bf{k}}(i\Delta_{+}e^{-i\phi_{\rm{k}}}a^{\dag}_{\rm{k}+}a^{\dag}_{-\rm{k}+}-
    i\Delta_{-}e^{i\phi_{\rm{k}}}a^{\dag}_{\rm{k}-}a^{\dag}_{-\rm{k}-})
    \  + h.c.\:,
\ee where the two independent order parameters $\Delta_{+,-}$ are
real,  and corresponding to the pairing order parameters on the
two helicity bands. Transforming back to the usual spin basis, the
explicit singlet/triplet order parameters are then given by,
$\Delta_{s(p)}=(\Delta_{+}\pm\Delta_{-})/2$, respectively, which
then yields the relation in Eq.\ (\ref{DefDeltaPM}). It is
worthwhile comparing with the noncentrosymmetric superconductors
in which the presence of quadratic dispersion term $k^2/2m$ in the
normal state leads to appearance of \emph{both} helicity bands at
the Fermi level. It follows that those with same (opposite) signs
of $\Delta_{\pm}$ resemble the singlet $s$-wave (triplet $p$-wave)
superconductors.\cite{YipTopology} Only the triplet
superconductors are associated with topological states such as
Andreev bound or zero-energy vortex bound states, which can exist
near the topological defects. For superconducting surface Dirac
fermions considered here, the appearance of a \emph{single}
helicity band near the Fermi level is the key
difference.\cite{Alicea} All the pairing symmetries that we
consider possess topological states as long as the superconducting
gap is not closed by the Zeeman field or the decreasing chemical
potential. Nevertheless, the mixed-parity pairing symmetry cases
with same (opposite) signs of $\Delta_{\pm}$ still resemble the
pure $s$-wave ($p$-wave) case, manifested by the eigenvalue of
antilinear spectrum-reflecting operator $A$ as in
Fig.\ref{MixedPhase}.

In conclusion, we determined the zero-modes for a variety of
the Jackiw-Rossi-Dirac Hamiltonians representing pairing of electrons
on the surface of a topological insulator
in presence of superconducting gaps of different symmetries.

\begin{center}
{\bf Acknowledgment}
\end{center}

The work has been supported by NSC Taiwan (C.K.L.) and NSERC
Canada (I.F.H.). C.K.L. also thanks Dr. Sungkit Yip for drawing
Ref.\cite{FuKane} to his attention and the relevant discussions.




\begin{thebibliography}{plain}
\bibitem{footnote} The most general Hamiltonian with the antilinear
reflection symmetry of its spectrum can also contain the covariant
momentum coupled to gauge field by
$p_i\rightarrow{p_i-iA_i\Gamma_3\Gamma_5}$, besides the terms of
chemical potential and Zeeman field \cite{HerbutLu}.
\end{thebibliography}
\end{document}